# Challenging Disability and Interaction Norms in XR: Cooling Down the Empathy Machine in Waiting for Hands

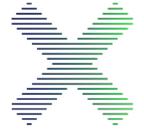

Virtual Reality (VR) is often hailed as the "ultimate empathy machine," framing disability as an experience to be simulated through such technologies—reducing disability to a spectacle of pity or inspiration. In response, we present *Waiting for Hands* (WfH), an interactive eXtended Reality (XR) installation that critiques this logic by: (1) Repurposing interaction norms in XR by building Alternative Controllers, and (2) Staging an absurd XR performance experienced using the built controllers to disrupt the sentimentalized disability narratives. The XR performance involves eight persons: two XR participants on stage and six seated audience members who watch a projected documentary about Hema Kumari, an Indian singer living with Rheumatoid Arthritis. The XR users' partially obscure the film, diverting audience attention through strange mouth and hand movements (instructed to perform in XR), creating a dual-layered experience that disrupts active viewership in Hema's story and introduces uncertainty. While XR is considered as a hot medium—fully immersive and sensory-dominant—this piece subverts that rationale by repurposing XR to produce absurdity and *alienation effect*. By defying empathetic, hence pitiable narratives of disability, we reflect: what ethical stance an XR performance can take to attune participants to a non-normative body schema while resisting spectacle?

## 1. Introduction: Motivation and Background

We are witnessing a significant transformation in how bodies interact with technologies in space, as the Internet overflows with videos of people wearing XR headsets in public—making "hand-gestures'' in mid-air while ordering at Starbucks, pinching fingers while crossing streets, and performing expansive 360-degree body movements to interact with these devices. XR (or eXtended Reality), an umbrella term for Virtual/Mixed Reality (VR/MR) technologies are proliferating, what Apple is referring to as the dawn of a new era in "spatial" computing.  Meta (formerly Facebook) calls this the "next evolution in social connection", *The Metaverse*, which the company claims, "will be built by everyone, […] by all sorts of imaginative people."  Yet, this vision of inclusivity excludes many, for example, our collaborator Hema

**Yesica Duarte**[+]
IT:U Interdisciplinary Transformation University Austria
yesicaduarte@gmail.com

**Puneet Jain**[+]
Concordia University, Montreal, Canada and Zürich University of the Arts, Switzerland
puneet.jain@zhdk.ch

Kumari, an Indian disabled artist, singer, and activist who lives with RA (Rheumatoid Arthritis). Hema's arthritis limits how her body can move in space and the hand gestures she can perform as any bodily movement is accompanied by elongated pain —particularly while using the hand-based XR controllers. Hema's inability to use XR technologies highlights how Big Tech companies reinforce ableist norms, informing what HCI researchers Gerling and Spiel (2021) call a "corporeal standard" in XR design—a template of a non-disabled body as the default user. Mott et al. (2020) likewise claim that the dominant XR devices are designed with "ability assumptions" (Wobbrock et al. 2011) hindering people with sensorimotor disabilities to access such technologies "when their abilities do not match these assumptions". Jain (2024) similarly critiques the implicit biases in XR technologies, demonstrating how they can be re-purposed, hacked and modified by drawing on lived experiences of disability.

In our proposed XR artwork, *Waiting for Hands* (WfH), we interrogate the interaction norms in XR alongside Hema by rejecting hand-based interaction as the default. Instead, we integrate Alternative Controllers[1] into the XR setup as a way to hack the very technologies under critique. The interface employed is a camera-based mouth and lip tracker (integrable in XR), co-developed with and for individuals with disabilities. This tool has been utilized by the authors in their previous artworks and is aligned with the same line of inquiry in *WfH*, as elaborated below.

In this case and context, the mouth is a key body part that Hema relies on for her everyday activities when her hands and legs are in tremendous pain. That is, her arthritis compels her to find new affordances—what disability scholar Arseli Dokumaci calls 'activist affordances': "possibilities of action that are almost too remote and therefore unlikely to be perceived and yet are perceived and actualized through great ingenuity and effort to ensure survival". (2023). In other words, "the tiny, everyday artful battles of disabled people for more livable worlds that otherwise remain unaccounted for." (Dokumaci 2023). These include Hema's act of using her teeth (instead of hands) to pull blankets over herself during winters, unscrewing water bottle caps by gripping them between her teeth and squeezing lemons by pressing them in her mouth. These tasks inform Hema story about how hand movements encompass the possibility of pain, and how the situated intelligence of the body shifts toward the mouth—all the acts that motivated the framework for the XR performance. Hence, leading to the design of Alternative Controllers, a mouth-based assistive interface to interact and navi-gate in XR, putting the mouth at the forefront in the XR performance, *Waiting for Hands* (WfH).

Concretely, our aim is to, firstly, expose the ableist biases embedded in the XR systems (i.e. revealing the mandatory use of hands in XR is an ableist assumption) and in return reshape the engagement of the audience in the performance (by shifting to mouth as a means of interaction but also distraction). The XR performance involves eight persons: two XR participants on stage and six seated audience members who watch a projected documentary about Hema. The XR users' partially obscure the film, diverting audience attention through strange mouth and hand movements (instructed to perform in XR, see Figure 1). That is, the audience watching Hema's story is not simply "transported" into her world; rather, they must contend with the disruptions caused by the alternative mouth-based interaction in XR. We create an open-ended and fragmented experience that requires the audience to actively negotiate meaning and reflect on what they are witnessing. By introducing friction through absurdity and the *alienation effect* (Verfremdungseffekt[2]) achieved in the performance design and setup, we "cool down" the "immersion" of these XR devices. We leverage assistive technologies to cultivate an alternative that centers the mouth and shifts the gaze away from objectifying Hema or portraying her disability as something to be pitied. Instead, the gazed is direct towards the assumptions of 'able-bodiedness' embedded in standard interaction design. However, it is worth noting that this project is not about making technology "more inclusive" in the way accessibility is often understood—adding accommodations as a gatekeeper to an existing paradigm. Instead, we propose rethinking the storytelling itself, one that enables reclamation for Hema and her life story of living with RA with the very [XR] technologies that marginalize them. Such stories, when told, as disabled queer femme of color educator-scholar Shayda Kafai argues "uncovers and gives voice to those who are unseen, marginalized, and forgotten," embracing "Disability Justice and its principle of collective liberation" (Kafai 2021). Taking this into consideration, our positionality is one of refusal—pushing back against the extractive logics of the voyeuristic gaze that reinforce ableist narratives. We want to resist perception of disability as spectacle, drawing inspiration from a rejection of the reductive framing of XR as an "empathy machine" (Milk 2015)—a concept that risks reinforcing disability as pitiable through empathy.

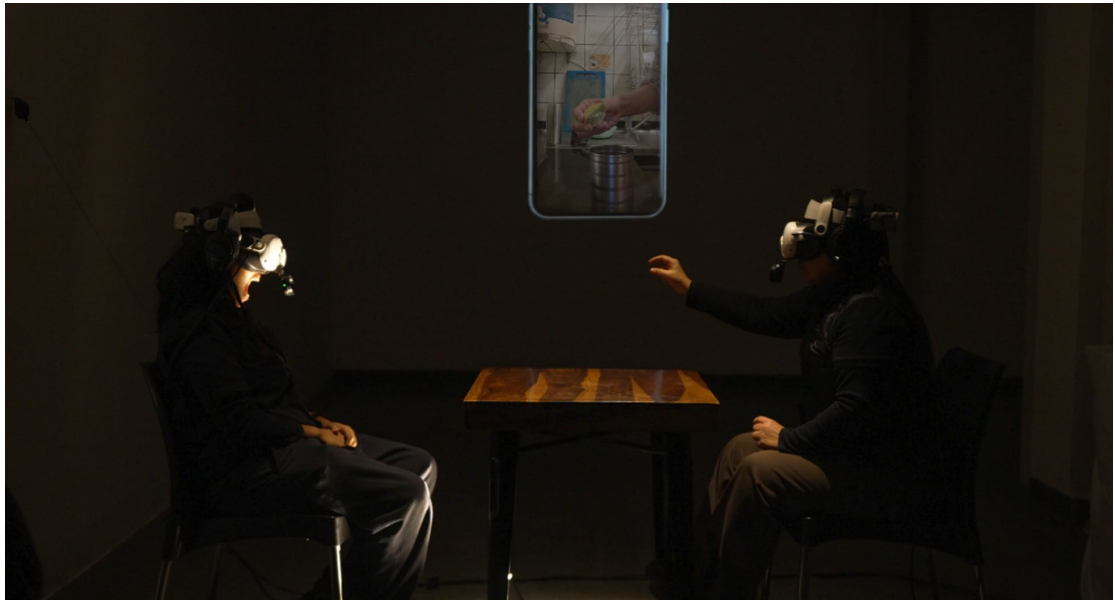

**Fig. 1.** Two participants on stage interacting with hands and mouth in XR.

Hema's Video Documentary plays behind them.

This proposal builds upon the prior experience of the authors in creating XR artworks and performances using Alternative Controllers as input modalities within XR systems. Concretely, *Waiting for Hands* draws on two research-creation projects: (1) First project explored a taxonomy of "disabled" mouth-gestures in XR detected by an assistive camera-based interface, developed with and for disabled artists and activists Eric Desrosiers and Christian Bayerlein (Jain and Salter 2025). These mouth gestures enabled the artists, Eric and Christian, to move and interact in VR/MR environments through chin, jaw, and tongue movements and were later realized as artistic strategies incorporated into the XR artwork *Crip Sensorama* (Jain, Bayerlein, and Duarte 2024) (2) The second project involved the development of custom wearable interfaces for *Pinch to Awaken XR* (Duarte and Rodriguez 2025), an XR performance that integrates DIY bio-sensing technologies with XR to explore body-technology relationships through the tracking of breathing patterns. A hybrid performer guides the audience through a digital/tangible performance, engaging them in a narrative that encourages synchronized breathing with a responsive environment, ultimately fostering bodily self-awareness. *WfH* extends this lineage of experimentation, proposing an interactive XR performance that not only questions dominant narratives around disability in XR but also reimagines how technology can attune to diverse bodies rather than demanding that bodies conform to technology.

## 2. Theoretical Framework

*WfH* explores how storytelling in XR can move beyond normative expectations toward a disability-centered perspective. We embrace Disability Justice as a framework that emphasizes with the idea of collective liberation. These concepts inform our design by shifting the focus away from a deficit model of disability and toward a more inclusive, resistance-based approach that centers disabled experiences in XR spaces. Our approach is inspired by what disability scholar Margaret Price (2024) calls *Crip spacetime*: "a material-discursive reality experienced by disabled people... whose meaning is made through the relations among words, bodies, objects, technologies, and environments." We aim to apply these ideas to reimagine spaces of XR by integrating nontraditional yet accessible interaction modalities through "assistive" XR—shifting away from ableist norms in spatial computing towards a more inclusive *Crip spacetime* framework for all.

Additionally, we also draw from radical theatre traditions to further inform our design approach: Brecht's Epic Theatre and Beckett's Theatre of the Absurd (Brecht 1964, Beckett 1953). Although these approaches differ in method and intention, both share an interest in questioning human existence and disrupting audience expectations. From Beckett, we inherit a dramaturgy of uncertainty, absurdity, and existential stasis. His plays—marked by repetition, silence, disjointed time, and non-sequiturs—reject linear progression and resist resolution, emphasizing instead the instability of meaning and the precarity of human experience. In *Waiting for Godot* (1953), for instance, time loops and non-action foreground the absurdities of waiting and dependency, conditions often pathologized within ableist frameworks. From Brecht, we draw on the *Verfremdungseffekt* or alienation effect, a technique designed to break the illusion of realism and encourage audiences to question what they see rather than passively consume it. Brecht sought to politicize theatre by making spectators aware of its constructed nature—through direct address, visible stagecraft, and episodic structure. *WfH* applies this principle by making the XR interfaces itself visible and "strange", using alternative or assistive controllers.

Extending these principles into XR, we engage with McLuhan's (1964) concept of hot and cool media, in relation to the level of engagement and interpretation those demand from the audience. XR typically functions as a hot medium—fully immersive, dictating the sensory experience, and fostering total absorption. Moreover, the reference of XR as "ultimate empathy machines" (Ventura et al. 2020, Hassan 2020), as the technologies that can enable one "to step into others' shoes" through simulations, has popularized these technologies as tools that can foster "positive attitudes" towards disabled communities (Chowdhury et al. 2019). This technological affordance has led to a growing trend of XR simulations designed to "feel" disability, such as experiences of blindness (Silverman 2015), schizophrenia (Penn,

118

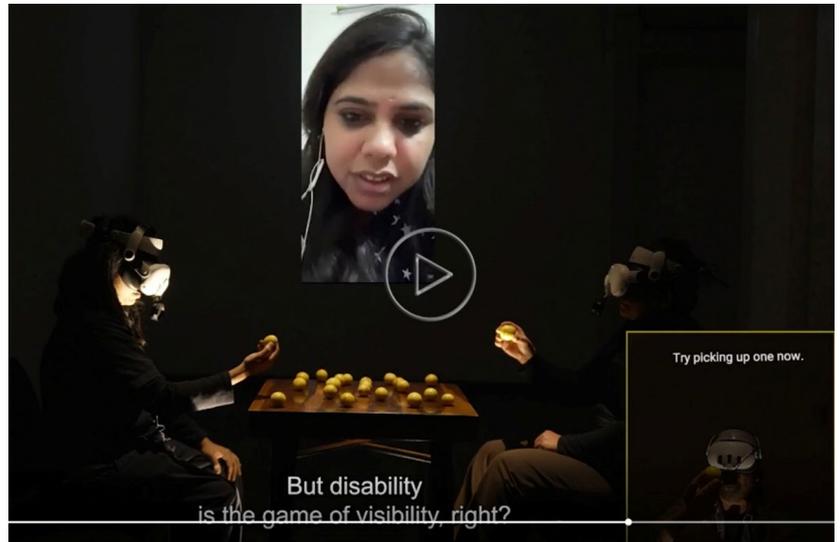

**Fig. 2.** Video documentation of the premiere exhibition at Khoj.

(https://tinyurl.com/wfh-videodocumentation).

Ivory, and Judge 2010), and even migraine (Excedrin 2023). Although these simulations claim, "to foster empathy and increase disability awareness, they often fall short of these goals, reinforcing harmful ableist attitudes." (Nario-Redmond and Gospodinov 2024).

This approach is problematic and has been especially critiqued in disability studies, due to simplifying complex social issues into temporary states and its tendency "to emphasize impairment-related deficits and dependence rather than the social, economic, and environmental factors that contribute to the challenges disabled people face." (Ibid 2024). The risk is perpetuating the notion that disability is merely a deficit to be overcome (medical model), rather than a diverse mode of being in the world (social model). Through the convergence of *Crip spacetime*, radical theatre, and media theory, we disrupt XR empathy machine to attune audience bodies to a shared narrative of disability culture and lived experience—one that resists spectacle, challenges normative embodiment, and reimagines interactive spaces.

## 3. Performance Description

Exhibit as a work-in-progress, *WfH* is an 8-minute performance, an interactive experience for eight participants: six seated audience members in a default theatrical setup, while two members are invited onto the stage, each wearing a XR headset adapted with a mouth and face tracker. The experience differs for each group.

The seated audience watches a video projection on a stage wall (see Figure 1), entering the world of Hema, a disabled artist based in Delhi, India. Meanwhile, the two XR participants, seated between the projection and the audience, face each other across a table. Immersed in an XR experience, they interact using their hands and mouth—first separately, then collaboratively, and ultimately engage with the seat-ed audience through physical objects. The performance is structured into two acts, separated by a two-minute interval (See Figure 2 and Link for Video Documentation). In Act One, the video introduces Hema's life and artistic practice. The seated audience watches as her story unfolds, while the two XR participants experience a scene centered around lemons. Initially, XR participants use their hands to pick floating virtual lemons, but as the experience progresses, their interactions shift to mouth-based controls, navigating a surreal highway, surrounded by lemons. As the first half of the documentary ends, a plaque announces the interval, and the curtains close.

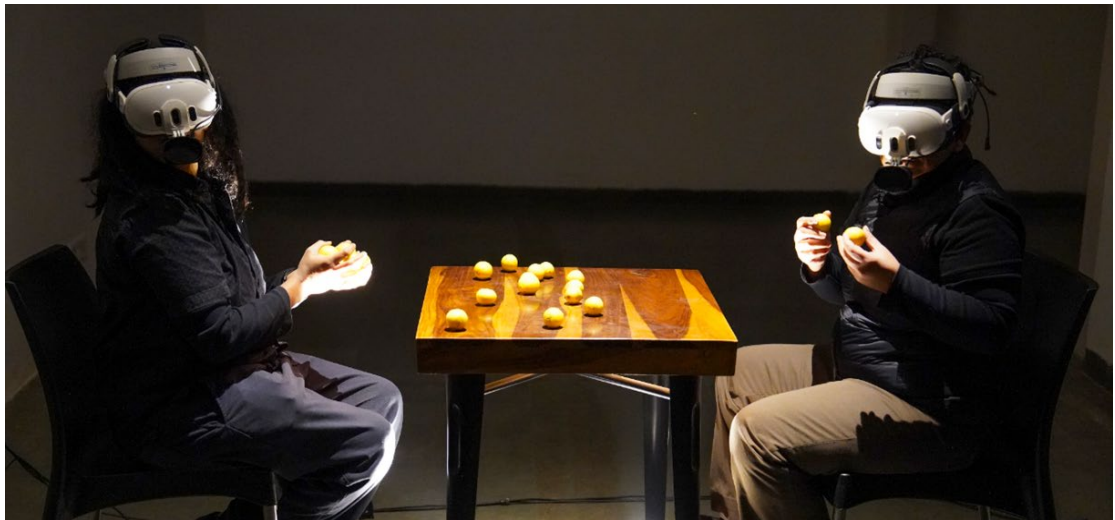

**Fig. 3.** XR participants on stage holding lemons while watching the seated audience in passthrough.

Act Two begins when the curtains reopen, revealing the table now covered with real lemons (see Figure 3). Hema reappears on screen, directly addressing the seated audience about her disability. Inside XR, the participants' interaction also shifts—this time, they pick up tangible, material lemons. The passthrough function (RBG cameras in the XR devices allowing users to see the material world through it) activates mixed reality, allowing them to see each other. Following a series of guided instructions, XR participants engage with the lemons in unexpected ways, eventually extending their actions to the seated audience (see Figure 4). As the seated audience members interact with XR participants, Hema appears on the screen singing a Bollywood folk song and the performance reaches its conclusion.

This setup allows us to explore not only new possibilities for storytelling of disability life and culture but also the potential for an artwork that hacks and disrupts contemporary VR/MR technologies. Through this absurd performance, we steer the audience away from affective immersion that reduces disability to simplistic emotional responses. The two XR participants disrupt the dominant paradigm of hand-based interaction in XR by using their mouths as primary interfaces, exposing how XR technologies assume certain bodily norms.

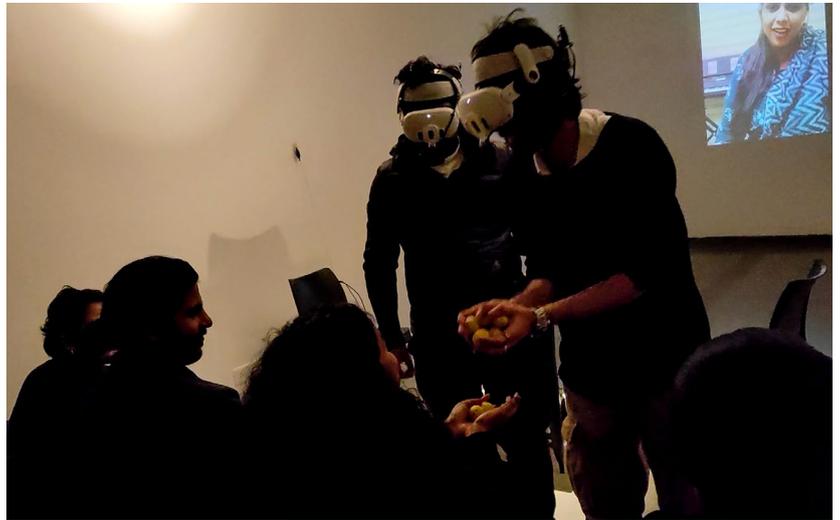

**Fig. 4.** Participants in XR interacting with the members of the audience, offering them lemons.

Within the performance, the XR devices function as Brechtian "distancing devices", disrupting default expectations and transforming interaction into a mechanism of critical reflection. This shift forces the audience to navigate between the two narratives: what is being shown (Hema's story) and what is being enacted (the XR interaction). This intentional displacement resists the objectification of the disability tourism, creating friction that prompts questions in the audience: What should we focus on? Should we empathize with Hema's story? Should they decode the XR interaction? How do these elements connect? What are the two participants doing? What are we watching? Why are they throwing material 'real' lemons to the audience?

These "unreliable actors" are active agents, which provoke and mislead by mediating possible meanings and distorting perspective. At times, reinterpreting Hema's gestures from the video, connecting it with XR participants actions—creating an ambiguous relationship between the visual and the embodied. This acts as a distancing mechanism, challenging conventional audience expectations and resisting the sentimentalization of disabled bodies. *WfH*, thus poses the following key question: How can an XR performance challenge the power dynamics embedded in current human-machine interfaces? What ethical stance can such a performance adopt to attune participants' senses to a non-normative body schema while resisting the voyeuristic gaze and the sentimentalization of disability? In turn, how can this attunement expose the deep-rooted prejudices shaping our perceptions of bodies, space, and XR technology?

## 4. Conclusion

*Waiting for Hands* stands at the intersection of *Crip spacetime*, radical theatre, and media theory, using XR to attune audience bodies to a shared narrative of disability culture and lived experience—one that resists spectacle, challenges normative embodiment, and reimagi-nes interactive spaces. By foregrounding mouth gestures as primary inputs in XR, the performance disrupts the assumed hierarchies of bodily interaction in XR, expanding the boundaries of what we are accustomed to experiencing in immersive environments.

Beyond expanding interaction modalities, the work intervenes in broader XR discourses by critically engaging with the ableist assumptions embedded in spatial computing. It challenges the industry's prevailing reliance on hand-based interactions and the 'empathy machine' narrative, proposing instead an XR that fosters active negotiation of meaning rather than passive consumption of experience. Through its approach, *WfH* reveals how accessibility can be a generative force for innovation in immersive media, rather than an afterthought. By questioning who XR is designed for and how it encodes normative assumptions about bodies and abilities, this work calls for a radical reimagining of interaction design—one that embraces diversity. As the field of XR continues to evolve and devices proliferate, *WfH* demonstrates that accessibility is not an act of objectification but a reflective space—one that challenges assumptions and pushes the field toward more inclusive and innovative futures.

**Acknowledgements.** This work was developed during the artistic residency BODIES-MACHINES-PUBLICS, that took place between October and December 2024, at Khoj International Artists' Association, Delhi, India. It is a two-year collaborative initiative between NAVE (Chile), Khoj International Artists' Association (India), Immersive Arts Space/ZHdK (Zurich, Switzerland), and Kornhaus Forum (Bern, Switzerland), supported by Swiss Art Council Pro Helvetia. https://khojstudios.org/project/waiting-for-hands/.

**Notes**

**+** indicates equal contributions.

**1.** "Alt Ctrl" are physical input mechanisms for interactive media that are distinct from traditional console-affiliated handheld controllers or computer inputs such as mice and keyboards. (Shake That Button n.d.)

**2.** Verfremdungseffekt (or "alienation effect") is a theatrical technique that disrupts the audience's emotional immersion to promote critical reflection on the play's social and political themes (Brecht 1964).